\documentclass[12pt]{article}
\usepackage{epsfig}

\def\beq{\begin{equation}}
\def\eeq#1{\label{#1}\end{equation}}
\def\eeqn{\end{equation}}

\def\beqa{\begin{eqnarray}}
\def\eeqa#1{\label{#1}\end{eqnarray}}
\def\eeqan{\end{eqnarray}}

\let\bar=\overbar

\def\Dslash{\not{\hbox{\kern-4pt $D$}}}
\def\dslash{\not{\hbox{\kern-2pt $\del$}}}

\def\msb{{\bar{\ssstyle M \kern -1pt S}}}

\def\Title#1{\begin{center} {\Large {\bf #1} } \end{center}}

\begin{document}

\Title{Rare K Decays}

\begin{center}{\large \bf Contribution to the proceedings of HQL06,\\
Munich, October 16th-20th 2006}\end{center}

\bigskip\bigskip


\begin{raggedright}  

{\it Michael Arenton\index{Arenton, M.}\\
Physics Department\\
University of Virginia\\
P.O. Box 400714\\
Charlottesville, VA 22904, USA}
\bigskip\bigskip
\end{raggedright}

\section{Introduction}

  We review recent results on rare K decays from KTeV and NA-48. By rare 
decays we mean both those modes where the experiments are pushing the 
branching fraction measurements and limits to lower and lower values and
also small branching fraction modes where  experimental advances now
allow their study with relatively large statistics.

\section{The KTeV Experiment}

\begin{figure}[!htb]
\begin{center}
\epsfig{file=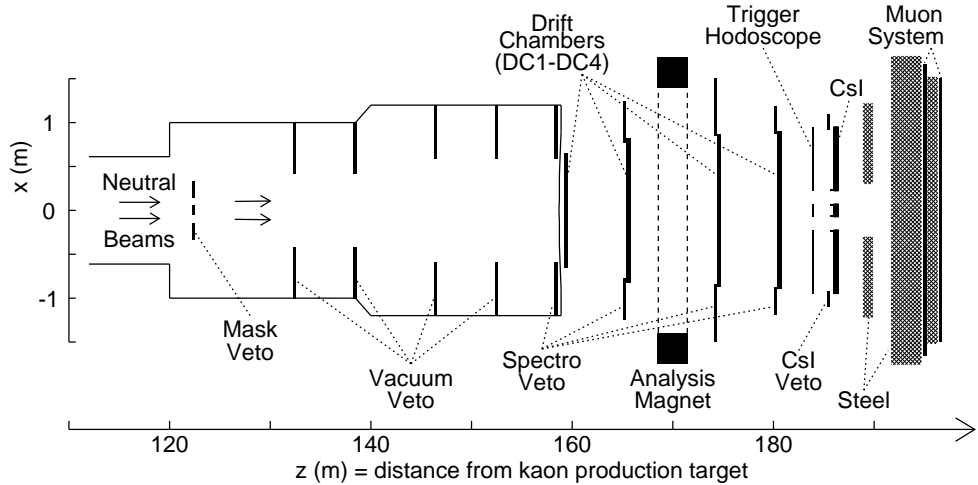,height=2.5in}
\caption{Plan view of KTeV E-832 configuration as used for rare decay
measurements (regenerator not shown).}
\label{fig:e832_layout}
\end{center}
\end{figure}

%
\begin{figure}[!htb]
\begin{center}
\epsfig{file=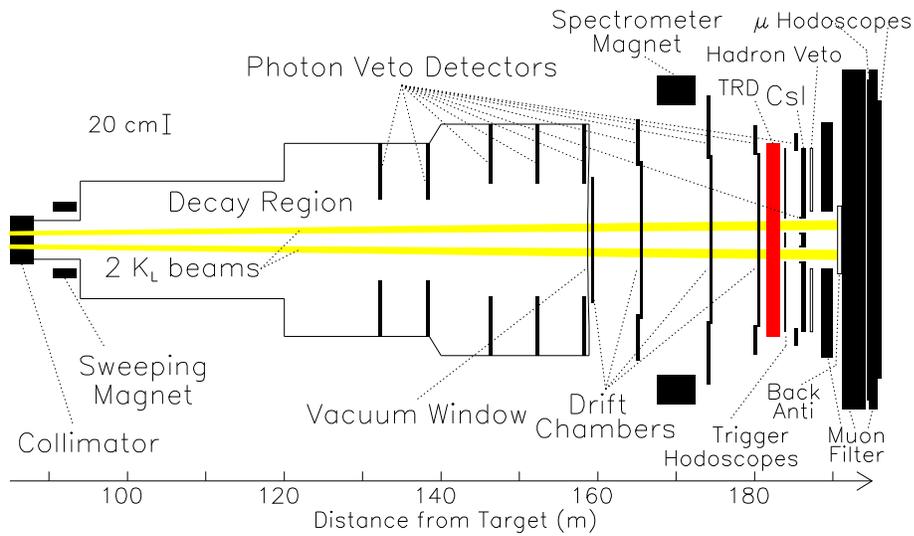,height=3.0in}
\caption{Plan view of KTeV E-799 configuration.}
\label{fig:e799_layout}
\end{center}
\end{figure}

Here we give a brief description of the KTeV 
experiment~\cite{Alavi-Harati:1999hd}~\cite{Alavi-Harati:1999zr}
. There were two
configurations, E832 and E799, shown respectively in Fig.~\ref{fig:e832_layout}
and Fig.~\ref{fig:e799_layout}.
. E832 was designed primarily to measure the
direct CP violation parameter $\epsilon'/\epsilon$, while
E799 was devoted to rare $K_L$ decays.
 Two $K_L$ beams were
generated by 800 GeV/c protons on a target. In E832 a regenerator in one
of the beams converted $K_L$ to $K_S$.  In E799 the regenerator was
removed to have two $K_L$ beams. The beams were run at higher intensity
in E799. The decays took place in a large vacuum decay region.

In both configurations a magnetic spectrometer consisting of two sets
of x and y drift chambers before and after an analysing magnet measured
charged particles.  Photons were measured in a 3100 element array of
pure CsI blocks 
which had energy resolution of $\sigma(E)/E = 0.45\% + 2\%/\sqrt(E)$. 
 This electromagnetc calorimeter was followed by layers
of steel and concrete absorber and scintillators for muon identification.
Several arrays of counters vetoed on the presence of charged particles or
photons outside the aperture of the spectrometer and calorimeter.
The E799 configuration also included a set of transition radiation
detectors for improved electron identification.

There were two data taking runs for each configuration, in 1997 and 1999.
The 1999 E832 run repeated the 1997 run with somewhat better running
conditions to check the systematics of the $\epsilon'/\epsilon$ measurement.
The 1999 E799 run was devoted to increasing the sensitivity for rare decays.
To this end the p$_t$ kick of the analysing magnet was reduced to 150 MeV/c
in 1999 from 200 MeV/c in 1997 to increase acceptance, particularly for
4 body decays. Also in 1999 several triggers were prescaled to increase
the data acquistion bandwidth for other triggers. E799 was sensitive to
2.5 X $10^{11}$ and 3.5 X $10^{11} K_L$ decays in 1997 and 1999 respectively.

\section{The decay $K_L \rightarrow \pi^+\pi^-\gamma$}

\begin{figure}[htb]
\begin{center}
\epsfig{file=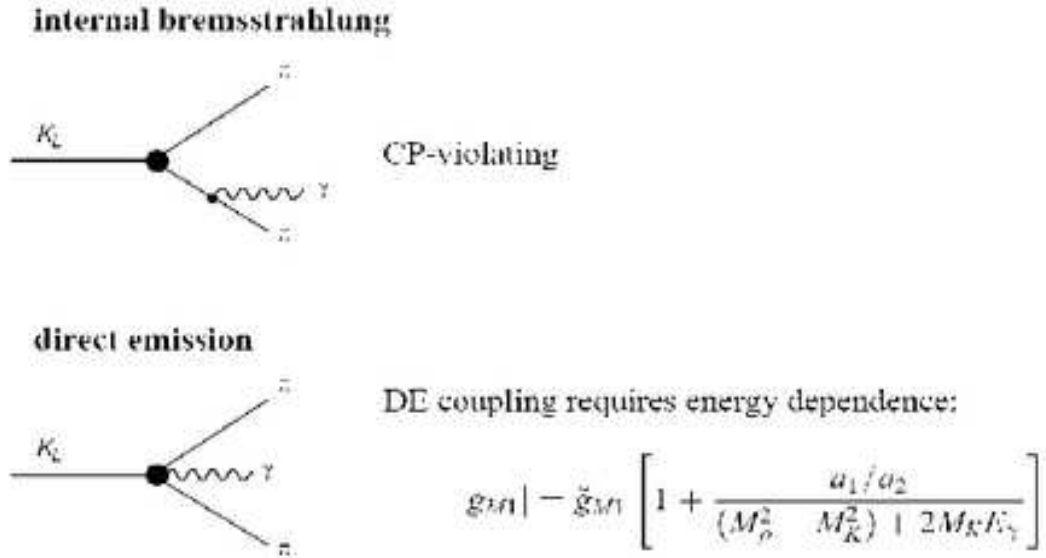,height=3.0in}
\caption{Amplitudes for the decay $K_L \rightarrow \pi^+\pi^-\gamma$ }
\label{fig:pipig_amplitudes}
\end{center}
\end{figure}

  The decay $K_L \rightarrow \pi^+\pi^-\gamma$ proceeds through the
amplitudes shown in Fig.~\ref{fig:pipig_amplitudes}. 
 The two main contributions,
of about equal magnitude, are from the CP violating Inner Bremsstralung (IB)
and the CP conserving Direct Emission terms.  The Direct Emission term is
mostly magnetic dipole (M1) radiation. This must be modified with a form factor
that is usually expressed in a $\rho$ pole form. As well as the M1 term there
might be an electric dipole (E1) term. The E1 term is especially interesting
because it is CP violating.
The experiments have now reached 
statistical levels where searching for the E1 term is possible.    

%

\begin{figure}[!htb]
\begin{center}
\epsfig{file=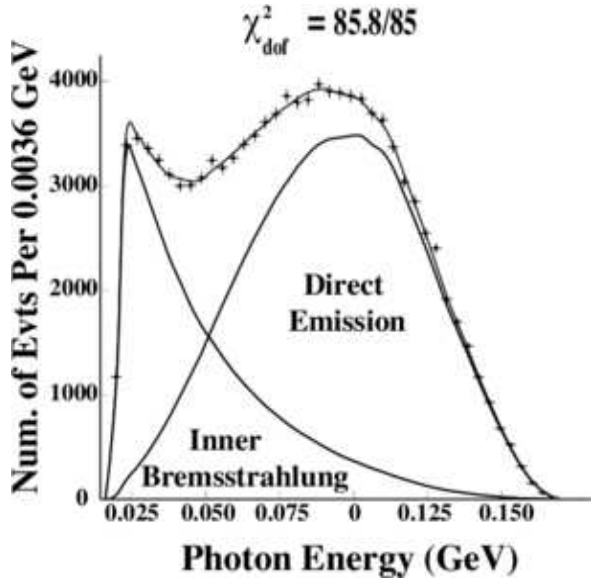,height=3.0in}
\caption{$E_{\gamma}$ distribution
 in $K_L \rightarrow \pi^+\pi^-\gamma$ with fit results.}
\label{fig:pipig_eg}
\end{center}
\end{figure}

 The KTeV results on $K_L \rightarrow \pi^+\pi^-\gamma$~\cite{Abouzaid:2006hy}
are based on a sample of 112,100 events 
over a background of 671 $\pm$ 41 events recorded in the 1997 run of E832.
The amplitudes are determined from a fit to the distribution of the $\gamma$
energy in the $K_L$ rest frame which shows a falling distribution at low
$E_{\gamma}$ from IB and a broad peak at high $E_{\gamma}$ from M1 DE. 
Interference of the M1 and E1 DE amplitudes would show up in the intermediate
energy region.  The distribution and fit are shown in figure 
\ref{fig:pipig_eg}. The DE
form factor parameters are found to be $g_{M1} = 1.198 \pm 0.035 \pm 0.086$
and $a_1/a_2 = -0.738 \pm 0.007 \pm 0.018 (GeV^2)$.  The ratio of the
direct emission to total decay rate is $0.0689 \pm 0.021$. An upper limit
for the magnitude of an E1 term is found to be $|g_{E1}| < 0.21$ at the
90\% confidence level.

%

%
\begin{figure}[!htb]
\begin{center}
\epsfig{file=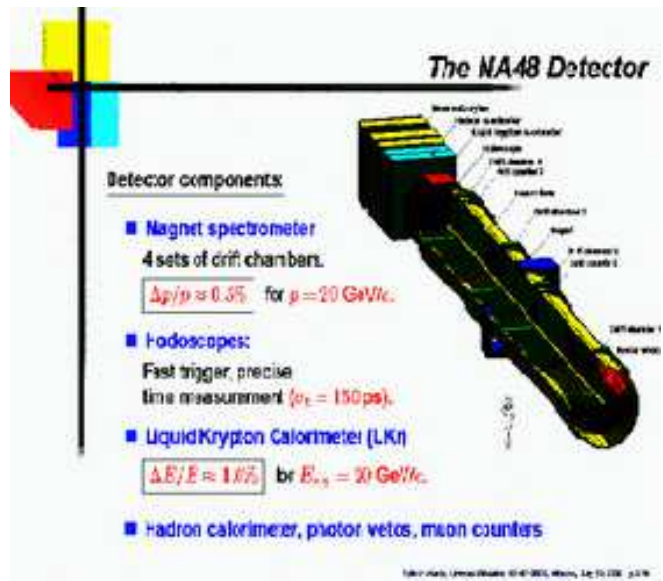,height=3.0in}
\caption{View of the NA48/2 detector. \protect\cite{Wanke:2006cy}}
\label{fig:na48_detector}
\end{center}
\end{figure}

%
\begin{figure}[!htb]
\begin{center}
\epsfig{file=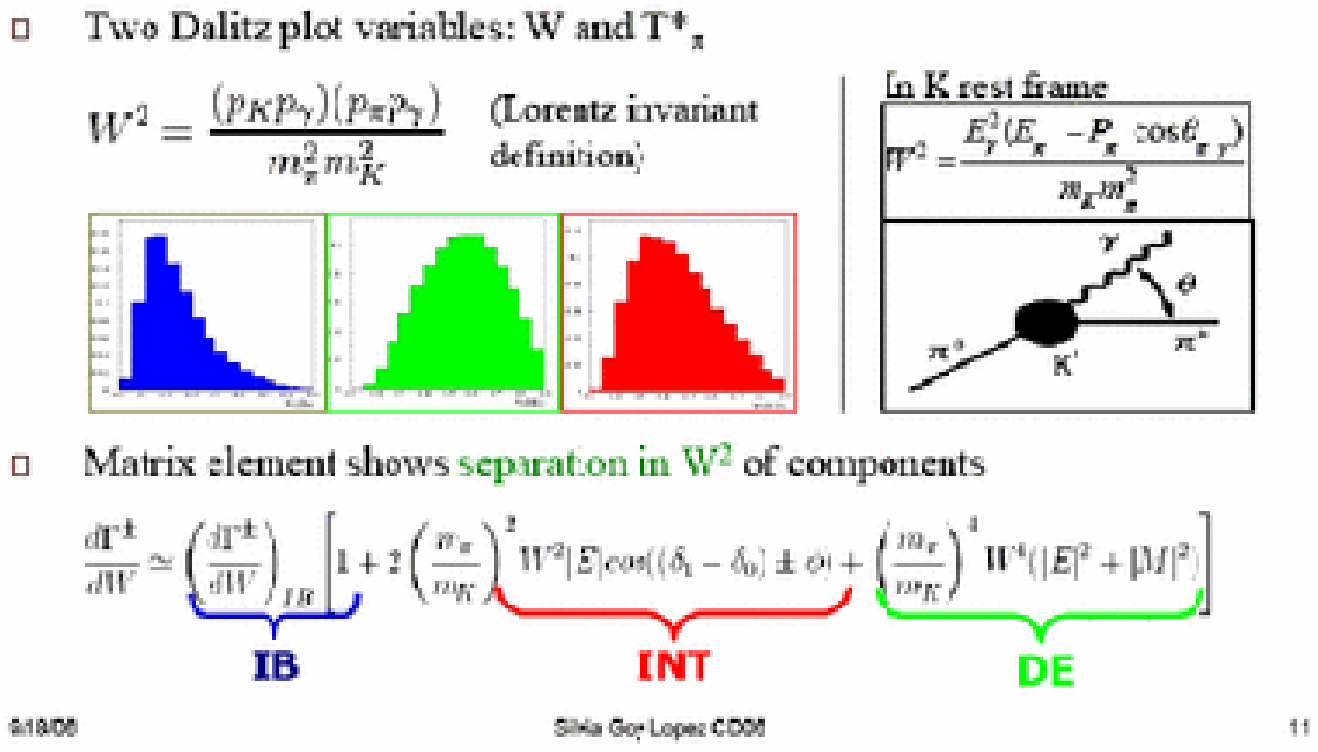,height=3.5in}
\caption{Variables describing the decay $K^\pm \rightarrow \pi^\pm\pi^0\gamma$ \protect\cite{GoyLopez:2006}}
\label{fig:na48_variables}
\end{center}
\end{figure}

KTeV has also searched for the E1 term in the related decay mode
$K_L \rightarrow \pi^+\pi^-e^+e^-$ \cite{Abouzaid:2005te}.
Along with amplitudes analogous to those of $K_L \rightarrow
\pi^+\pi^-\gamma$ this decay also has a term related to the K charge
radius.  The angular distributions of the $\pi\pi$ and ee pairs 
provide additonal information making this decay more sensitive to
E1 contributions.  We find an upper limit of
$|g_{E1}|/|g_{M1}| < 0.04$ at 90\% confidence level.

\section{NA48 results on $K^\pm \rightarrow \pi^\pm\pi^0\gamma$}

 The charged kaon extension of the NA48 experiment, NA48/2, has recently
presented results on the related radiative decay 
$K^\pm \rightarrow \pi^\pm\pi^0\gamma$. Like
$K_L \rightarrow \pi^+\pi^-\gamma$ this decay has IB and DE
amplitudes, but here the IB amplitude is CP conserving and much larger
than DE.

  The NA48/2 detector is shown in Fig.~\ref{fig:na48_detector}
 \cite{Wanke:2006cy}  The NA48/2 experiment was primarily directed to
searching for direct CP violation in $K^{\pm} \rightarrow 3\pi$ decays.
It used simultaneous $K^{\pm}$ beams of 60 $\pm$ 3 GeV/c.

The Dalitz plot variables $W^2$ and $T^*_\pi$ used to describe the 
$K^\pm \rightarrow \pi^\pm\pi^0\gamma$ decay are shown in 
Fig.~\ref{fig:na48_variables}. One searches for the interference term
between the IB and the electric dipole direct emission. Previous 
experiments have not seen evidence of this interference. This analysis
was based on 124,000 events, which is 30\% of the available data and
5 times the statistics of previous experiments.

In the analysis it was necessary to solve two problems that might cause
distortions in the $W^2$ distribution that could mimic direct emission
or interference terms. The first is  misassignment of the $\gamma$'s, which
was reduced by cuts on the $\pi^0$ and $K^{\pm}$ masses and requirement
of agreement of the charged and neutral vertices. The second is
backgrounds from the $K^{\pm} \rightarrow \pi^{\pm}\pi^0$ and
$K^{\pm} \rightarrow \pi^{\pm}\pi^0\pi^0$ decays with coalesed $\gamma$'s,
which were reduced using techniques to split coalesed $\gamma$'s.
In the final analysis
 these backgrounds were reduced to less than 1\% of the direct
emission level.

%
\begin{figure}[!htb]
\begin{center}
\epsfig{file=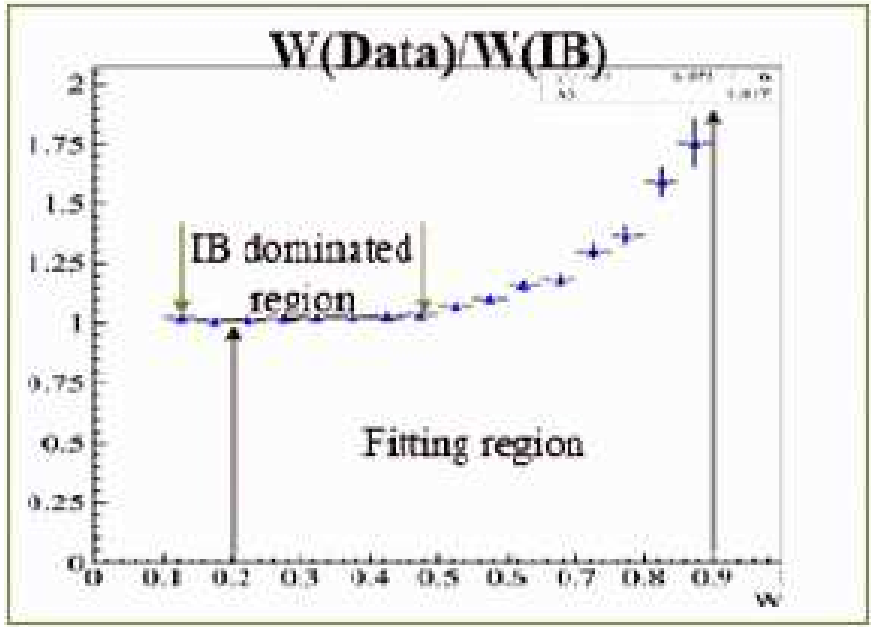,height=3.5in}
\caption{W distribution of NA48/2 data divided by Inner Bremmstralung shape.\protect\cite{GoyLopez:2006} }
\label{fig:na48_wfit}
\end{center}
\end{figure}

Fig.~{\ref{fig:na48_wfit} shows the measured W distribution divided by that
expected for IB alone. Clearly this deviates from unity at high W.
A fit yields a preliminary result for the fraction of direct emission of 
$(3.35 \pm 0.35_{stat} \pm 0.25_{syst})\%$ and of the interference
$(-2.67 \pm 0.81_{stat} \pm 0.73_{syst})\%$. 
These values are highly correlated, with a correlation coefficient of
-0.92.  This result is the first observation of the electric dipole 
interference term with high statistical certainty.

\section{$K_L \rightarrow \pi^+\pi^-\pi^0\gamma$ and $K_L \rightarrow
\pi^+\pi^-\pi^0e^+e^-$}

 First results on the radiative decays $K_L \rightarrow \pi^+\pi^-\pi^0\gamma$
and  $K_L \rightarrow \pi^+\pi^-\pi^0e^+e^-$ have been obtained by KTeV.
$K_L \rightarrow \pi^+\pi^-\pi^0\gamma$ is expected to be dominated by the
inner bremmstralung process with a theoretical branching fraction of
$(1.65 \pm 0.03) \times 10^{-4}$ for $E_{\gamma} < 10 MeV$
\cite{D'Ambrosio:1996zy}. The direct emission contribution is expected
to be very small: $BR|_{direct} = {(8a_1 + a_2 -10a_3)^2}{^.}2^.10^{-10}$
where the $a_i$ are unknown parameters of order 1 \cite{Ecker:1993cq}.
For $K_L \rightarrow \pi^+\pi^-\pi^0e^+e^-$ there are no published
theories.  There should be IB and DE terms similar to
 $K_L \rightarrow \pi^+\pi^-\pi^0\gamma$ with virtual photon conversion to 
an $e^+e^-$ pair. In addition there should be a charge radius amplitude.

KTeV has observed $K_L \rightarrow \pi^+\pi^-\pi^0\gamma$ both in data from
E832 with $\pi^0 \rightarrow \gamma\gamma$ and from E799 with
$\pi^0 \rightarrow e^+e^-\gamma$ yielding signals of 2853 and 2847
events respectively, as shown in Fig. \ref{fig:pipipizero_gamma}.
  A preliminary result for
the branching ratio with $E^{cm}_{\gamma} > 10 MeV$ is
$BR = (1.70 \pm 0.03_{stat} \pm 0.04_{syst} \pm 0.03_{ext syst})\times 10^{-4}$
in good agreement with theory.

%
\begin{figure}[htb]
\begin{center}
\epsfig{file=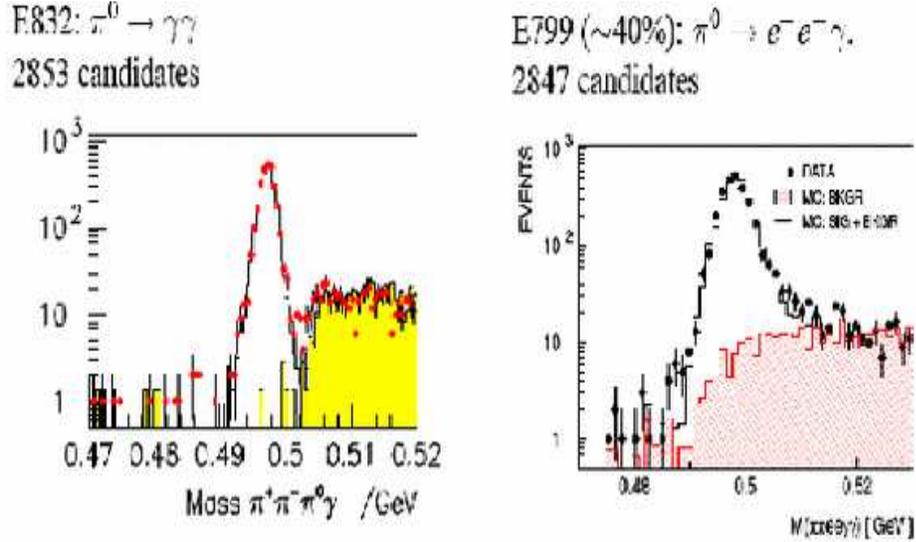,height=3.0in,width=5.0in}
\caption{$\pi^+\pi^-\pi^0\gamma$ and $\pi^+\pi^-\pi^0_D\gamma$
mass distributions from E832 and E799 data. }
\label{fig:pipipizero_gamma}
\end{center}
\end{figure}

KTeV has made a first observation of $K_L \rightarrow \pi^+\pi^-\pi^0e^+e^-$
in the E799 data. In 40\% of the data 132 candidates are observed with
an estimated background level of $1.2 \pm 0.9$ event. This is shown in
Fig. \ref{fig:pipipizero_ee}.
 The preliminary result for $E_{ee} > 20 MeV$ is
$BR = (1.60 \pm 0.18_{stat})\times 10^{-7}$.

%
\begin{figure}[htb]
\begin{center}
\epsfig{file=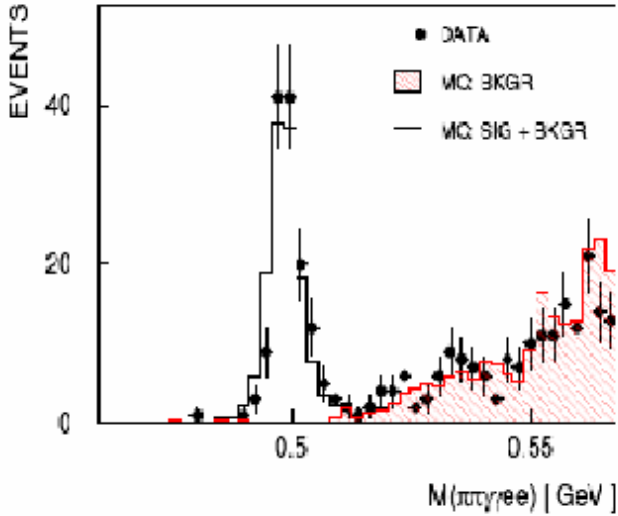,height=3.0in}
\caption{$\pi^+\pi^-\pi^0_De^+e^-$
mass distribution from E799 data. }
\label{fig:pipipizero_ee}
\end{center}
\end{figure}

\section{$K_L \rightarrow e^+e^-\gamma$}

Radiative decays of the type $K_L \rightarrow \gamma^{(*)}\gamma^{(*)}$
are of interest largely because of their role in the measurement of the
CKM matrix element $|V_{td}|$ from the decay $K_L \rightarrow \mu^+\mu^-$.
This decay proceeds partly from a short distance coupling related to
 $|V_{td}|$, but also has a long distance couplling related to 
$K_L \rightarrow \gamma^{(*)}\gamma^{(*)}$ that must be subtracted.

The $K_L \rightarrow \gamma^{(*)}\gamma^{(*)}$ decays have been
described by two form factor models. One is the vector dominance inspired
model of Bergstr\"om, Masso and Singer (BMS) \cite{Bergstrom:1983rj}.
The other is the chiral perturbation theory model of D'Ambrosio, Isidori and
Portoles (DIP) \cite{D'Ambrosio:1997jp}.
The parameters of these models can be determined by fits to the $m_{ee}$
distribution of the data. The BMS model contains a parameter $\alpha_{K^*}$.
However experiments actually determine the quantity $C\alpha_{K^*}$
where
 $C = (8\pi\alpha_{em})^{1/2}G_{NL}f_{K^*K\gamma}m_{\rho}^2/(f_{K^*}f_{\rho}^2A_{\gamma\gamma})$.
A number of experiments using various decay modes have presented results for
$\alpha_{K^*}$ but are inconsistent because the values of the parameters making
up C have changed over time. Therefore KTeV chooses to quote
 $C\alpha_{K^*}$ and
compare it to $C\alpha_{K^*}$ from other experiments.

The KTeV form factor measurements are based on a sample of 83,000
$K_L \rightarrow e^+e^-\gamma$ decays.  Of particular importance in the
form factor fits is the handling of radiative corrections.  KTeV has
developed a Monte Carlo including the complete set of second order
radiative diagrams.

The corrected preliminary KTeV results are a branching fraction of
$(9.25 \pm 0.03_{stat} \pm 0.07_{syst} \pm 0.26_{ext syst})\times 10^{-6}$,
$C\alpha_{K^*} = -0.517 \pm 0.030_{fit} \pm 0.022_{syst}$ for the BMS model
and $\alpha_{DIP} = -1.729 \pm 0.043_{fit} \pm 0.028_{syst}$ for the
DIP model. A comparison of values of $C\alpha_{K^*}$ from various
decay modes is shown in Fig.\ref{fig:calpha}.

%
\begin{figure}[htb]
\begin{center}
\epsfig{file=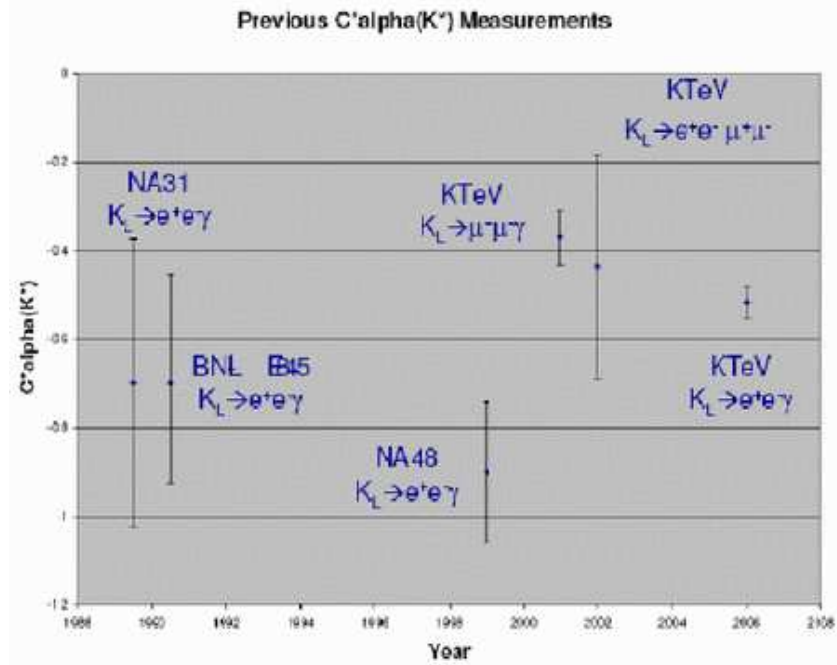,height=3.5in}
\caption{Results on $C\alpha_{K^*}$ from various decay modes }
\label{fig:calpha}
\end{center}
\end{figure}

\section{$K_L \rightarrow \pi^{\pm}e^{\mp}{\nu}e^+e^-$}

KTeV has made the first measurements of the decay
$K_L \rightarrow \pi^{\pm}e^{\mp}{\nu}e^+e^-$
 \cite{Kotera}.
This is of course related to the radiative $K_{e3}$ decay mode
$K_L \rightarrow \pi^{\pm}e^{\mp}\nu\gamma$. New tests of chiral
perturbation theory are enabled by these measurements.

Because of the missing $\nu$ there are less kinematic constraints
available to use in signal selection.  The worst backgrounds come from
$K_L \rightarrow \pi^+\pi^-\pi^0$ with $\pi^0$ decays to $e^+e^-\gamma$
or $e^+e^-e^+e^-$, $K_L \rightarrow \pi^{\pm}e^{\mp}\nu\pi^0$ with
$\pi^0 \rightarrow e^+e^-\gamma$, and
 $K_L \rightarrow \pi^{\pm}e^{\mp}\nu\gamma$ where the $\gamma$ converts
in material in the spectrometer.  The analysis relies heavily on the 
full identification power of the CsI Calorimeter and the TRD's.

A sample of 19466 candidate events is obtained with a background of
about 5\% from about 25\% of the E799 data.  A preliminary result for
the branching fraction of $K_L \rightarrow \pi^{\pm}e^{\mp}{\nu}e^+e^-$
with $m_{e^+e^-} > 0.005$GeV and $E_{e^+e^-} > 0.03$GeV is
$(1.281 \pm 0.010_{stat} \pm 0.019_{syst} \pm 0.035_{ext syst})\times 10^{-5}$.

A theoretical calculation in chiral perturbation theory \cite{tsuji} has
been made of the quantity 
$R = \Gamma(K_L \rightarrow \pi^{\pm}e^{\mp}{\nu}e^+e^-, m_{ee} >~0.005 GeV)
/\Gamma(K_L \rightarrow \pi^{\pm}e^{\mp}\nu)$.
At leading order in the theory the predicted value of R is
$4.06 \times 10^{-5}$ whereas at next to leading order ($p^4$) it is
$4.29 \times 10^{-5}$.  The experimental result corresponds to
R = $(4.54 \pm 0.15) \times 10^{-5}$ which is $3.2\sigma$ from the
leading order and $1.7\sigma$ from the next to leading order calculation.

%
\begin{figure}[!htb]
\begin{center}
\epsfig{file=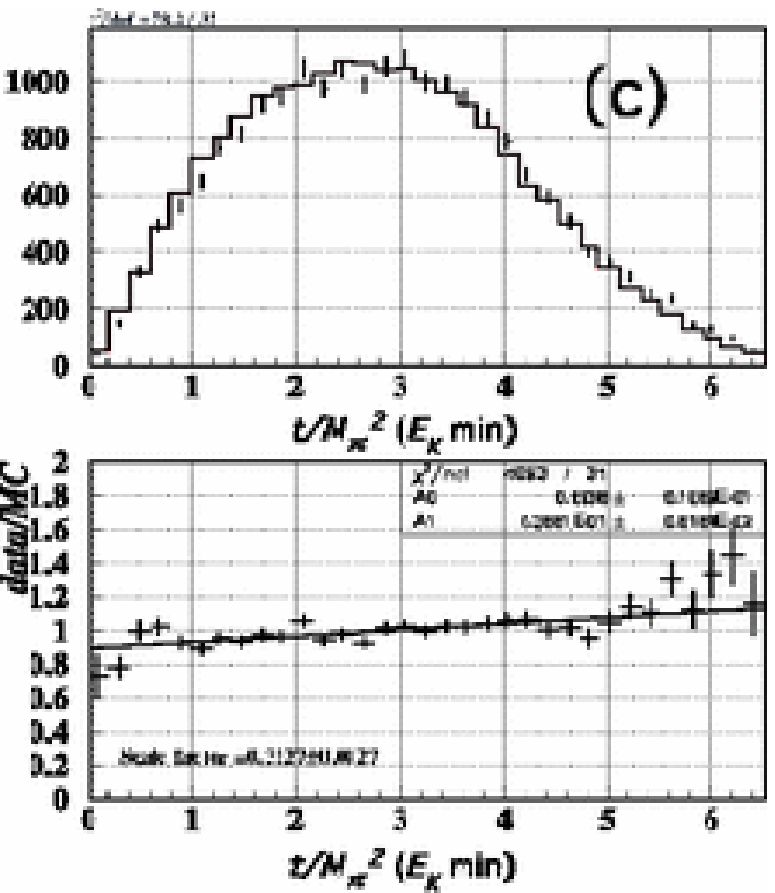,height=3.5in}
\caption{Distribution of the minimum solution for t. Upper plot shows\
data (points) and leading order $\chi$PT calculation (lines). Lower plot\
show the ratio of data/calcuation. }
\label{fig:t_mn_lo}
\end{center}
\end{figure}

One may also examine agreement with theoretical predictions by looking
at the distributions of the kinematics of the decay.  One such variable is
the momentum transfer t. Because of the missing $\nu$, experimentally there
are two possible solutions for t in each event. Figs. \ref{fig:t_mn_lo} 
and \ref{fig:t_mn_nlo} show the distributions of the minimum solution
for t. The points in both of these figures are the same. 
Fig. \ref{fig:t_mn_lo} shows a comparison  to the distribution calculated
with leading order chiral perturbation theory while Fig. \ref{fig:t_mn_nlo}
shows the comparison to next to leading order theory. One sees a better
agreement with next to leading order, as shown in the ratio plots in the
bottom parts of the figures, the ratio being flatter for NLO than for LO.
The same conclusion is reached when the maximum solution for t is 
examined (not shown).

%
\begin{figure}[!htb]
\begin{center}
\epsfig{file=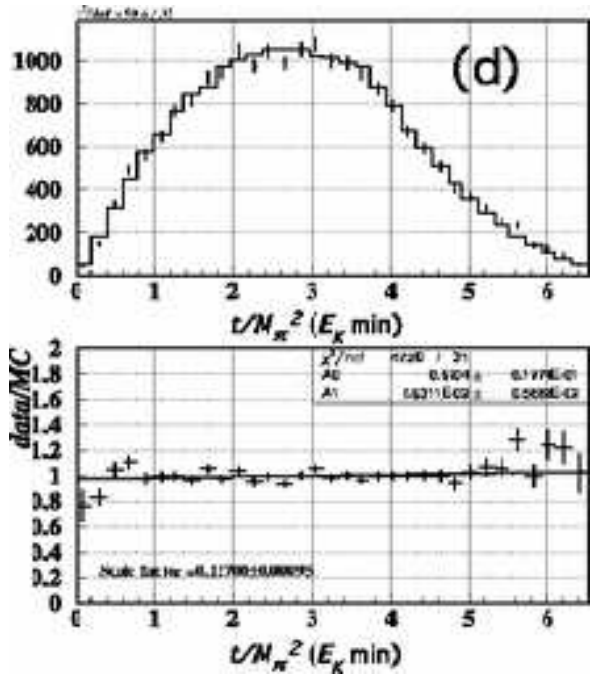,height=3.5in}
\caption{Distribution of the minimum solution for t. Upper plot shows data(points) and next to leading order $\chi$PT calculation (lines). Lower plots shows the ratio of data/calculation. }
\label{fig:t_mn_nlo}
\end{center}
\end{figure}

\section{$\pi^0 \rightarrow e^+e^-$}

$K_L$ decays are a copious source of ``tagged'' $\pi^0$'s.  KTeV has used
these to measure the rare decay $\pi^0 \rightarrow e^+e^-$.  To lowest order
this is described by the diagram shown in Fig. \ref{fig:pizero_diagram}
\cite{drell}. Various calculations based on vector dominance or 
chiral perturbation theory
 predict branching fractions somewhat higher
than this unitarity limit \cite{bergstrom_two}~\cite{savage}~\cite{ametller}
~\cite{gomez}~\cite{knecht}.

%
\begin{figure}[!htb]
\begin{center}
\epsfig{file=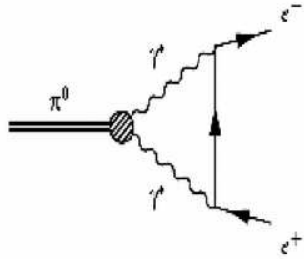,height=1.5in}
\caption{Lowest order diagram describing $\pi^0 \rightarrow e^+e^-$. }
\label{fig:pizero_diagram}
\end{center}
\end{figure}

%
\begin{figure}[!htb]
\begin{center}
\epsfig{file=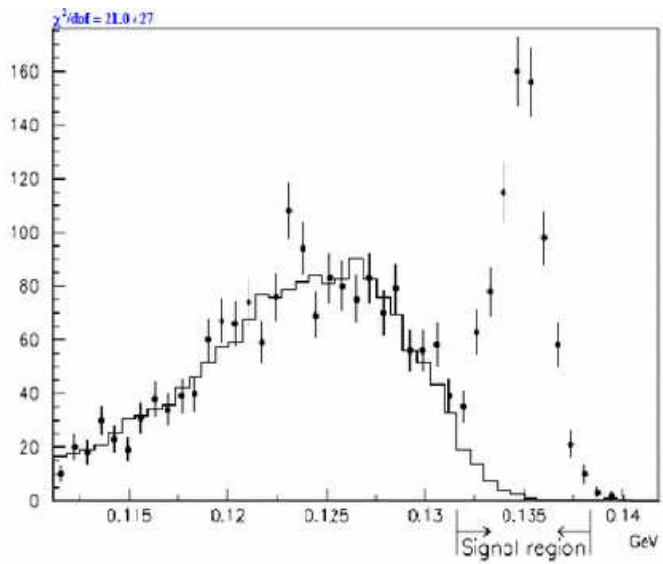,height=3.0in}
\caption{$m_{ee}$ distribution after all other cuts. }
\label{fig:pizero_mee}
\end{center}
\end{figure}

%
\begin{figure}[!htb]
\begin{center}
\epsfig{file=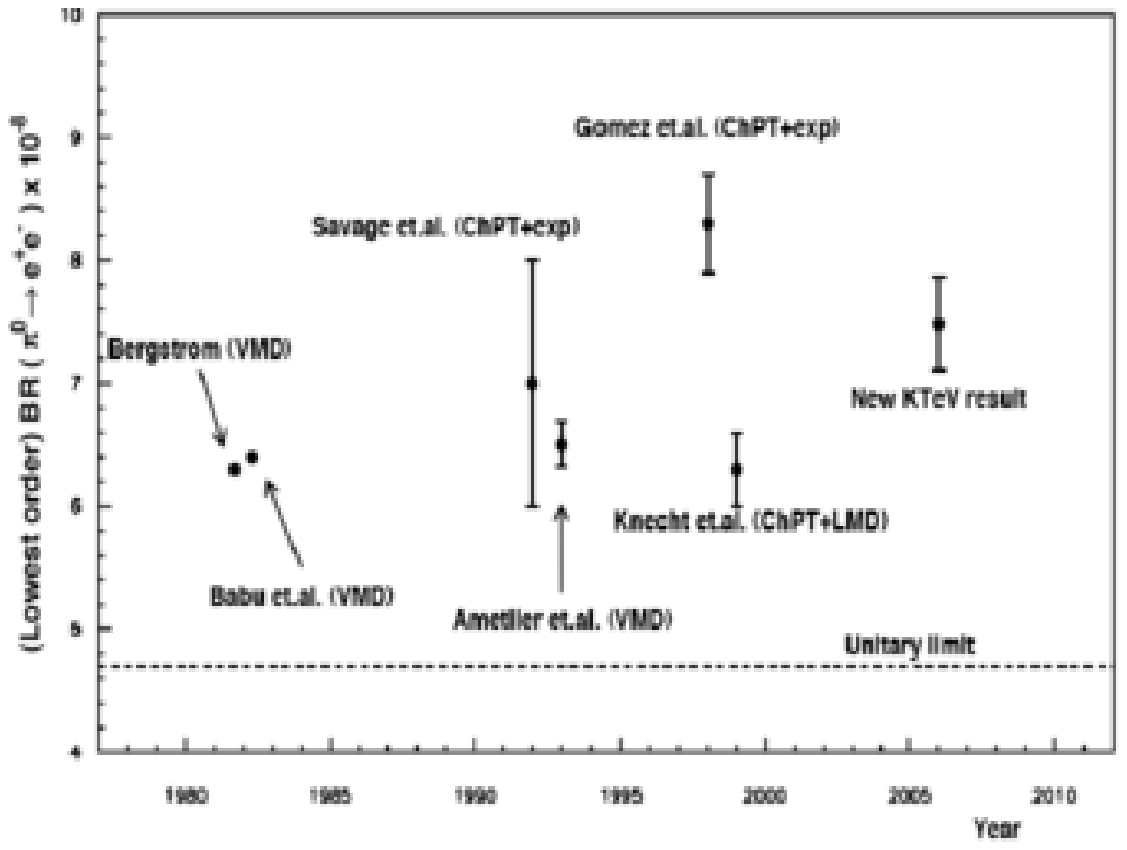,height=3.0in}
\caption{Comparison of the measured $\pi^0 \rightarrow e^+e^-$ branching
fraction to theories of refs \protect\cite{bergstrom_two} to \protect\cite{knecht}. }
\label{fig:pi0ee_results}
\end{center}
\end{figure}

After appropriate analysis cuts the $m_{ee}$ distribution shown in
Fig.\ref{fig:pizero_mee} is obtained, for events of the final state
$\gamma\gamma\gamma\gamma e^+e^-$ with two $\gamma\gamma$ pairs
consistent with $\pi^0$'s and $m_{\gamma\gamma\gamma\gamma ee}$
consistent with $m_K$. The peak has 794 events with a background of
$53.2 \pm 9.5$ events. The branching fraction for $\pi^0 \rightarrow
e^+e^-$ with $x>0.95$ is $(6.56 \pm 0.26_{stat} \pm 0.10_{syst}
 \pm 0.19_{ext syst}) \times 10^{-8}$, where $x = m_{ee}/m_{\pi^0}$.

Fig. \ref{fig:pi0ee_results} shows this result in comparison with the
unitarity limit and various theoretical calculations. This measurement
is $7\sigma$ above the unitarity limit.

\section{Searches for Lepton Flavor Violating Decays}

KTeV has searched for the lepton flavor violating decays
$K_L \rightarrow \pi^0\mu^{\pm}e^{\mp}$,
$K_L \rightarrow \pi^0\pi^0\mu^{\pm}e^{\mp}$, and
$\pi^0 \rightarrow \mu^{\pm}e^{\mp}$. There is nothing new to
report on $K_L \rightarrow \pi^0\mu^{\pm}e^{\mp}$ but recently new
preliminary results have been obtained on the latter two modes.
(The $\pi^0$ decay analysis is in effect a subset of the
$K_L$ analysis.)  

In the past, analyses of this sort have been done by defining a ``box''
in some kinematical space (usually in a 2 dimensional plot of the mass 
of all the particles making up the decay versus a transverse momentum that
should be zero for a true decay), not looking at the data in the box while
cuts to reduce backgrounds are established by studying their effects on
the events near to, but outside the box, and finally opening the box
and comparing the number of events found to what is expected from
background evaluations. This is not in general the most sensitive
procedure however, since the true signal would not usually be evenly
distributed over a rectangular box.  

Instead KTeV has based its analysis on a probability distribution function
(PDF) formed from the distributions of mass and $p_t^2$.  The PDF
distribution for the decay mode $K_L \rightarrow \pi^0\mu^{\pm}e^{\mp}$
is shown in Fig. \ref{fig:pdf_region} which illustrates the search regions.
The PDF distribution for $K_L \rightarrow \pi^0\pi^0\mu^{\pm}e^{\mp}$
is very similar.  

%
\begin{figure}[htb]
\begin{center}
\epsfig{file=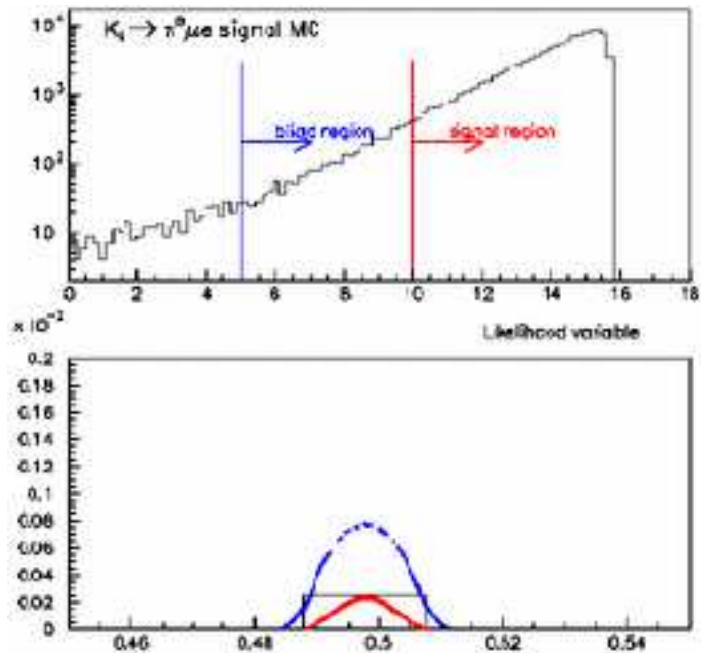,height=3.5in}
\caption{Signal probability distribution function for $K_L \rightarrow \pi^0\mu e$.  }
\label{fig:pdf_region}
\end{center}
\end{figure}

Determination of the background is a key to this analysis.  Initial
Monte Carlo calculations indicated that backgrounds originate from
several common decay modes.  Because of this and the very small
branching fractions being probed it was not practical to calculate
background levels by Monte Carlo simulations.  Instead the data itself
was used, loosening some cuts so that the PDF distributions of the
background could be examined and then rescaling by the effects of the
final cuts. Figure \ref{fig:pdf_fit}} shows this.
The result of this is predicted backgrounds of $0.44 \pm 0.12$ events 
of $K_L \rightarrow \pi^0\pi^0\mu^{\pm}e^{\mp}$ and
$0.03 \pm 0.02$ events for $\pi^0 \rightarrow \mu^{\pm}e^{\mp}$.

%
\begin{figure}[htb]
\begin{center}
\epsfig{file=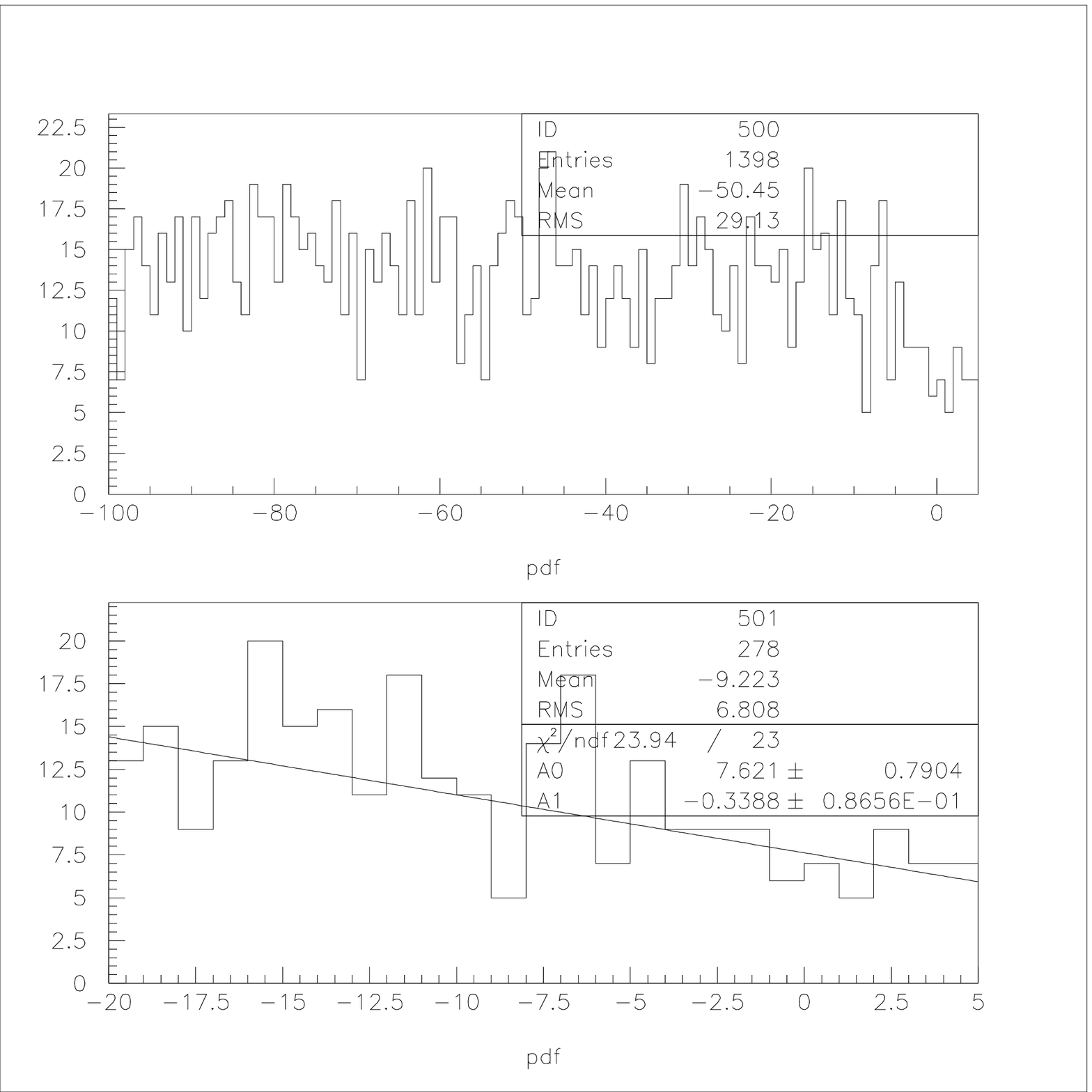,height=3.5in}
\caption{Signal probability distribution distributions for relaxed cut background samples. Note that the signal region is PDF $>$ 10. }
\label{fig:pdf_fit}
\end{center}
\end{figure}

When the blind region in the PDF distribution was examined no events
were found.
Using the Feldman-Cousins method with the predicted backgrounds and
no events seen yields preliminary 90\% confidence level upper limits
of $BR(K_L \rightarrow \pi^0\pi^0\mu^{\pm}e^{\mp}) < 1.58 \times 10^{-10}$
and $BR(\pi^0 \rightarrow \mu^{\pm}e^{\mp}) < 3.63 \times 10^{-10}$.

\section{Conclusion}

Sensitive results have recently been obtained on a number of rare K decays
\cite{echeu}.
In $K_L \rightarrow \pi^+\pi^-\gamma$ and $K_L \rightarrow 
\pi^+\pi^-e^+e^-$  accurate measurement of the inner bremsstrahlung and M1
direct emission components have been made and searches done for the E1
amplitude.  In the analogous charged decay 
$K^{\pm} \rightarrow \pi^{\pm}\pi^0\gamma$ NA48/2 has found the first
clear evidence for the E1 interference term.  KTeV has observed the
decays $K_L \rightarrow \pi^+\pi^-\pi^0\gamma$ and
$K_L \rightarrow \pi^+\pi^-\pi^0e^+e^-$.  KTeV has made high statistics
measurements of $K_L \rightarrow e^+e^-\gamma$.  KTeV has made the
first measurements of $K_L \rightarrow \pi^{\pm}e^{\mp}{\nu}e^+e^-$.
KTeV has made accurate measurements of $\pi^0 \rightarrow e^+e^-$.
Finally KTeV has new results on searches for the lepton flavor violating
decays $K_L \rightarrow \pi^0\pi^0\mu^{\pm}e^{\mp}$ and
$\pi^0 \rightarrow \mu^{\pm}e^{\mp}$.

\bigskip

I thank my colleagues on KTeV and Augusto Ceccucci of NA48 for discussions.
This work was supported by the U.S. Department of Energy


\begin{thebibliography}{99}
\bibitem{echeu}
  KTeV results from other decay modes have been presented at this conference
 by E.~Cheu.
\end{thebibliography}
\end{document}